\documentclass[conference]{IEEEtran}
\IEEEoverridecommandlockouts
\usepackage{cite}
\usepackage{amsmath,graphicx,url,times,booktabs,tabularx,amsfonts,siunitx}
\usepackage{algorithmic}
\usepackage{graphicx}
\usepackage{textcomp}
\usepackage{xcolor}
\usepackage{latexsym}
\usepackage{bbding}
\usepackage{url}
\usepackage{multirow}
\usepackage{siunitx}
\usepackage{microtype}
\usepackage{soul}
\usepackage{color}
\usepackage{footnote}
\usepackage[colorlinks,linkcolor=blue,]{hyperref}

\def\BibTeX{{\rm B\kern-.05em{\sc i\kern-.025em b}\kern-.08em
    T\kern-.1667em\lower.7ex\hbox{E}\kern-.125emX}}
\begin{document}

\let\OLDthebibliography\thebibliography
\renewcommand\thebibliography[1]{
  \OLDthebibliography{#1}
  \setlength{\parskip}{0pt}
  \setlength{\itemsep}{5.1pt plus 0.1ex}
}

\title{Leveraging Pre-trained AudioLDM for Sound Generation: A Benchmark Study}

\author{\IEEEauthorblockN{1\textsuperscript{st} Given Name Surname}
\IEEEauthorblockA{\textit{dept. name of organization (of Aff.)} \\
\textit{name of organization (of Aff.)}\\
City, Country \\}}

\author{
      \IEEEauthorblockN{
      Yi Yuan$^{1*\thanks{* Equal contributions}}$,
      Haohe Liu$^{1*}$,
      Jinhua Liang$^{2}$,
      Xubo Liu$^{1}$,
      Mark D. Plumbley$^{1}$,
      Wenwu Wang$^{1}$,
     }
      \\
      \IEEEauthorblockN{$^1$Centre for Vision, Speech and Signal Processing (CVSSP), University of Surrey, UK}
      \IEEEauthorblockN{$^2$Centre for Digital Music (C4DM), Queen Mary University of London}
}

\maketitle

\begin{abstract}
Deep neural networks have recently achieved breakthroughs in sound generation. Despite the outstanding sample quality, current sound generation models face issues on small-scale datasets (e.g., overfitting), significantly limiting performance. In this paper, we make the first attempt to investigate the benefits of pre-training on sound generation with AudioLDM, the cutting-edge model for audio generation, as the backbone. Our study demonstrates the advantages of the pre-trained AudioLDM, especially in data-scarcity scenarios. In addition, the baselines and evaluation protocol for sound generation systems are not consistent enough to compare different studies directly. Aiming to facilitate further study on sound generation tasks, we benchmark the sound generation task on various frequently-used datasets. We hope our results on transfer learning and benchmarks can provide references for further research on conditional sound generation. 
\end{abstract}


\section{Introduction} 
The development of deep learning models has led to a surge of interest in sound generation. Different strategies have been developed for sound generation tasks with input contents as diverse as tag~\cite{Liu-tts}, text~\cite{audiogen, tan2022naturalspeech} and video~\cite{specvqgan}. Sound generation systems are useful tools for content creation in applications such as virtual reality, movies, music, and digital media~\cite{liu2022separate, movie, Liuvocoder}. 

Recently, significant progress has been made in high-fidelity text-to-sound generation~\cite{diffsound,audiogen,audioldm}. Such sound generation systems are usually data-hungry to train. For example, AudioGen~\cite{audiogen} collected ten different datasets for training. However, this is not viable in some real-world applications, (e.g., animal sound and environmental sound generation), where the collection and labelling work for this specific domain is a time-consuming and costly process, leading to limited datasets scale in practice. How to overcome the data scarcity issue is a significant challenge in sound generation research. Several methods are adapted to tackle this issue. Rongjie \textit{et al.}~\cite{makeaudio} introduced a pseudo prompt enhancement approach to increase the data quantity, while the quality of new augmented data is unstable. Given these considerations, it is intuitive to ask: \textit{can we find an effective solution to train a sound generative model with a small-scale dataset?}

Studies have shown that pre-trained models can achieve faster and more accurate adaptions in tasks with limited data~\cite{pre-training1,pre-training2}. Concretely, pre-trained models are neural networks that have already trained on a massive corpus and can be fine-tuned into downstream tasks. Over the last few years, pre-training strategies have achieved enormous success across multiple fields including text ~\cite{textgeneration,Liucaption,Bert}, image 
~\cite{imagegeneration1,imagegeneration2} and audio ~\cite{classifier1,Liuretrival,liuaudio2text}. However, the effectiveness of a pre-trained model for sound generation is an under-explored topic. This paper takes the first step on investigating the effectiveness and feasibility of pre-training in text-to-sound generation with AudioLDM~\cite{audioldm}, the state-of-the-art audio generation model. Our results show that pre-trained models can achieve better performance on sound generation, especially for small-scale datasets. 

Besides, previous sound generation studies used different methodologies for evaluation, making it difficult for us to evaluate the model performance fairly. Without a set of constant metrics, researchers may find it hard to reproduce the results of other models. Aiming to provide an efficient and reliable reference for further sound generation research, this paper introduces a new benchmark with pre-trained AudioLDM on four commonly used audio datasets: AudioCaps~\cite{audiocaps}, AudioSet~\cite{audioset}, Urbansound8K~(US8K)\cite{urbansound8k} and ESC50~\cite{esc50}. 
Furthermore, our new benchmark contains most of the evaluation metrics applied in previous works~\cite{audiogen,diffsound,audioldm}, including Fréchet Distance~(FD), Inception Score~(IS)\cite{isc}, Fréchet Audio Distance~(FAD)~\cite{fad} and Kullback-Leibler~(KL) divergence. With several qualitative experiments, we provide insights into the effectiveness of these metrics in evaluating sounds. Our contributions are as follows: 

\begin{itemize}
    \item We showcase that transferring the pre-trained AudioLDM is beneficial for sound-generation tasks in both sample quality and training efficiency, especially for small-scale datasets.
    \item We benchmark the sound generation task by presenting the result of AudioLDM with multiple settings on four commonly used sound datasets.
\end{itemize}

\section{Related Work}

\noindent
\textbf{Conditional sound generation.}
Kong \textit{et al.}~\cite{sampleRNN} took the first step on conditional generation by taking labels as input and generating waveforms with recurrent neural network~(RNN). Then, Liu \textit{et al.}~\cite{Liu-tts} tried to synthesise sound with latent discrete features obtained from a vector quantised-variational autoencoder~(VQ-VAE)~\cite{vqvae} in the frequency domain~(e.g. mel-spectrogram). By compressing the mel-spectrogram into a sequence of tokens, the model can generate sounds with long-range dependencies. Recently, remarkable progress has been made in text-to-sound generations. Diffsound~\cite{diffsound} explores generating audio with a diffusion-based text encoder, a VQ-VAE-based decoder and a generative adversarial network (GAN)-based vocoder. Taking texts as input, Diffsound utilized a contrastive language image pre-training~(CLIP) model~\cite{clip} for text embedding before sending it to the encoder. To alleviate the scarcity of text-audio pairs, they proposed a text-generating strategy by combining mask tokens and sound labels. 
AudioGen~\cite{audiogen} used a similar encoder-decoder structure to Diffsound~\cite{diffsound}, while generating waveform directly instead of using a vocoder. They used a transformer-based encoder to generate discrete tokens and a pre-trained Transfer Text-to-Text Transformer~(T5)~\cite{t5} for text embedding. To increase the quantity of sound, they mixed audio samples at various signal-to-noise ratios~(SNR) and collect $10$ large datasets. 

\noindent
\textbf{Evaluation metrics for sound generation.}
Since subjective metrics for sound-generating systems usually require a huge amount of time and workload, various objective metrics were applied for this task. However, previous works often adopted different evaluation metrics, which makes it difficult to compare them in a common ground. Kong \textit{et al.}~\cite{sampleRNN} used Inception Score~\cite{isc} as the criterion. Liu \textit{et al.}~\cite{Liu-tts} trained a sound classifier to verify the sample quality. Diffsound~\cite{diffsound} applied Fréchet Inception Distance~(FID)~\cite{fid} and Kullback-Leibler~(KL) divergence to compute the sample fidelity, as well as a pre-trained audio caption transformer~(ACT) to calculate a sound-caption-based loss. AudioGen~\cite{audiogen} evaluated the result with KL divergence and Fréchet Audio Distance~(FAD). 

\section{Method and Datasets}

\subsection{AudioLDM}
Given text-based information (i.e. a written description containing single or multiple sound events), the objective of text-to-sound generation is to generate an audio clip that presents correct sound events as the text description. 

Our experiments are carried out with AudioLDM~\cite{audioldm}, a continuous latent diffusion-based model~(LDMs) for text-to-sound generations. Inspired by previous text-to-sound models, AudioLDM adapts a similar encoder, decoder, and vocoder architecture. By comparison, the text encoder in previous studies~\cite{diffsound, audiogen, makeaudio} is replaced by a Contrastive Language-Audio Pre-training~(CLAP) model. Specifically, the CLAP consists of two encoders, a text encoder $\textit{f}_{text}$ that extracts text description \textit{y} into text embedding $\boldsymbol{E}^{y}$ and an audio encoder $\textit{f}_{audio}$ that computes audio embedding $\boldsymbol{E}^{x}$ from audio samples \textit{x}. 
CLAP trains two encoders along with two projection layers using a symmetric cross-entropy loss, resulting in an aligned audio-text latent space. By utilizing the audio embedding during training and text embedding during sampling, AudioLDM can significantly reduce the demand for text-sound pairs and enable a self-supervised paradigm of LDM optimization. 
The latent diffusion model contains two processes: 1) a forward process that gradually transforms the data into a standard Gaussian distribution; and 2) a reverse process that generates data from the Gaussian distribution by denoising in reverse order as the forward process.
During the forward process, the continuous latent representation $\boldsymbol{z}_{0}$ from the mel-spectrogram is transformed into a standard Gaussian distribution $\boldsymbol{z}_{n}$ by gradually adding a scheduled Gaussian noise in \textit{N} steps. The transition probability of each time step \textit{n} is:
\begin{align}
q(\boldsymbol{z}_{n}|\boldsymbol{z}_{n-1})&=\mathcal{N}(\boldsymbol{z}_{n};\sqrt{1-\beta_{n}}\boldsymbol{z}_{n-1},\beta_{n}\boldsymbol{I}), \\
\label{forwardprocess}
q(\boldsymbol{z}_{n}|\boldsymbol{z}_{0})&=\mathcal  N(\boldsymbol{z}_{n};\sqrt{\bar{\alpha}_{n}}\boldsymbol{z}_{0},(1-\bar{\alpha}_{n})\boldsymbol{\epsilon}),
\end{align} 
where $\epsilon \sim \mathcal{N}(0,\textit{\textbf{I}})$ denotes the Gaussian noise with level ${\alpha}_{n}$ and schedule $\beta_{n}$.
The latent diffusion model is trained with the re-weighted training objective~\cite{DDPM,audioldm}, given by
\begin{align}
\label{trainingobjective}
L_{n}(\theta)=\mathbb{E}_{\boldsymbol{z}_{0},\boldsymbol{\epsilon},n}\left \| \boldsymbol{\epsilon} - \boldsymbol{\epsilon}_{\theta}(\boldsymbol{z}_{n},n,\boldsymbol{E}^{x}) \right\|^2_{2},
\end{align}
where $\theta$ denotes the trainable parameters in the LDMs. Benefits from the aligned audio-text space from CLAP, the reverse transition probability, $p_{\theta}(\boldsymbol{z}_{n-1}|\boldsymbol{z}_{n},\boldsymbol{E}^{y})$, can be parameterized by both $\boldsymbol{\epsilon}_{\theta}(\boldsymbol{z}_{n},n,\boldsymbol{E}^{y})$ and $\boldsymbol{\epsilon}_{\theta}(\boldsymbol{z}_{n},n,\boldsymbol{E}^{x})$~\cite{audioldm}. Data can be generated by performing reverse diffusion from a sample of standard Gaussian distribution with the reverse transition probability~\cite{DDPM}. We will compare the difference between conditioning with $\boldsymbol{E}^{x}$ and $\boldsymbol{E}^{y}$ in our experiment.

\subsection{Dataset}

\noindent
\textbf{Dataset for pre-training AudioLDM.} 
The datasets for per-taining AudioLDM include AudioSet~\cite{audioset}, AudioCaps~\cite{audiocaps}, Freesound\footnote{\url{https://freesound.org/}}, and BBC Sound Effect library~(BBC SFX)\footnote{\url{https://sound- effects.bbcrewind.co.uk/search}}. AudioSet~\cite{audioset} is the largest dataset with \num{527} text labels and around \num{5000} hours of sound. AudioCaps~\cite{audiocaps} is a smaller dataset with additional human-written audio captions. Both AudioCaps~\cite{audiocaps} and AudioSet~\cite{audioset} are captured from YouTube. FreeSound is a dataset provided by a public sound community with various durations. BBC SFX is a high-quality dataset from BBC with a wide range of sound effects. Note that most pre-training datasets originally come with text-sound pairs~(e.g. AudioCaps and BBC SFX), we only use audio embedding as the condition during the self-supervised training of LDMs. Combining all four datasets, we have totally \num{3.3}M ten-second sound clips to train our AudioLDM.

\noindent
\textbf{Dataset for benchmark study.}
To establish the baselines for the transferring study, we perform experiments on three common audio datasets with different volumes.

Two relatively small datasets we use are Urbansound8K~(US8K)\cite{urbansound8k} and ESC50~\cite{esc50}. 
US8K contains \num{8000} sound clips with \num{10} classes and ESC50 has \num{50} classes with only \num{40} samples for each class.
We randomly select 870 samples in US8K and 400 samples in ESC50 for evaluation.
Apart from ESC50 and Urbansound8K, we also perform experiments on AudioCaps \cite{audiocaps} to further enhance our study. AudioCaps contains around $47000$ ten-second audio data with more diverse sound events. Although AudioCaps is included in the pre-training dataset of AudioLDM, we find further fine-tuning on AudioCaps can improve model performance on the AudioCaps evaluation set.

\section{Evaluations and Experiments}

\subsection{Evaluation}
\begin{figure}[htbp]
    \centering
    \includegraphics[width=1.0\linewidth]{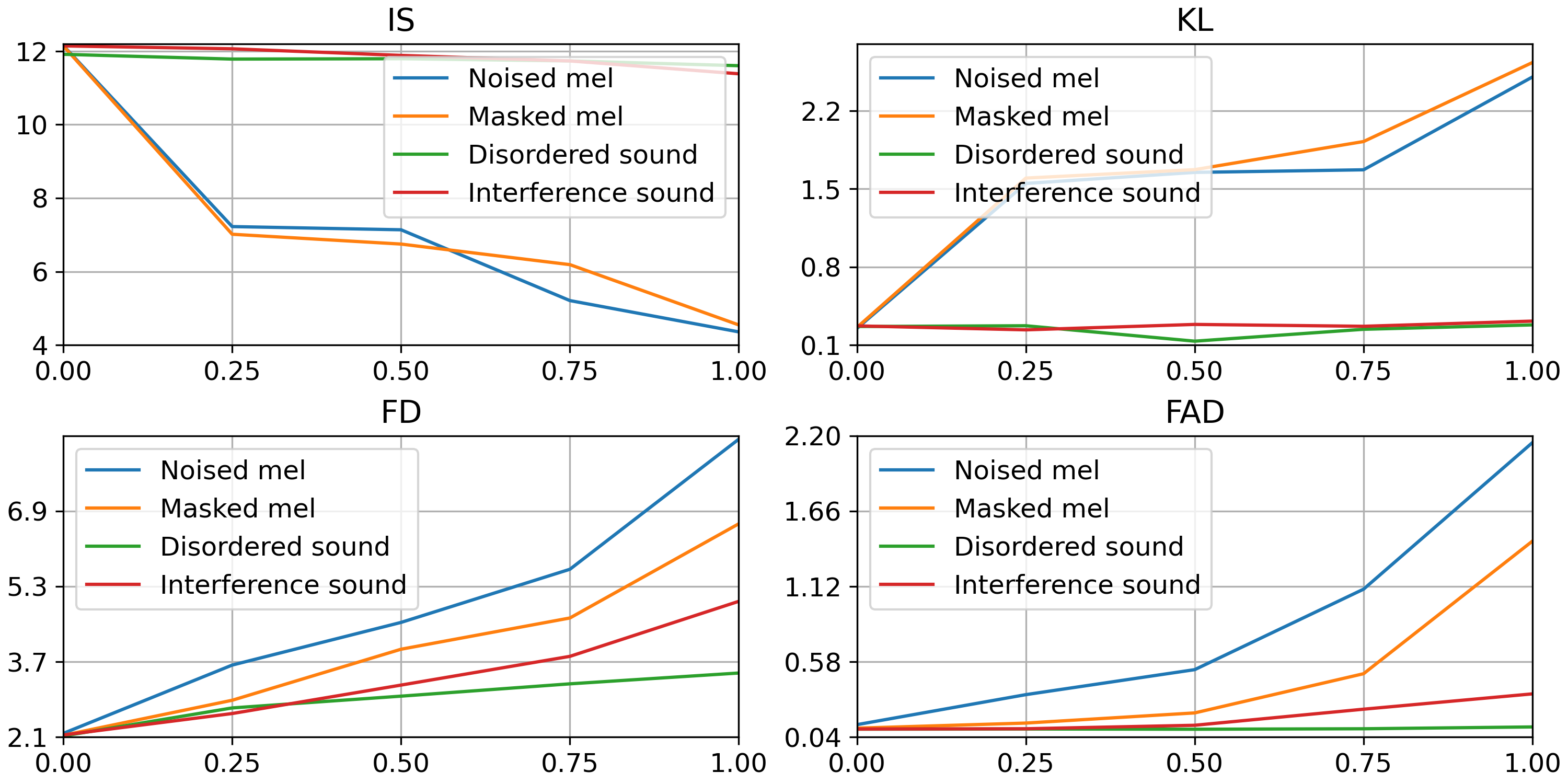}
    \caption{The metrics are evaluated with the increase of the percentage~(from 0 to 1) of pre-processed data on: 1) adding noise on the mel-spectrogram; 2) masking value on the mel-spectrogram; 3) making disorder sound events; 4) adding interfering sound events. Regarding metrics capabilities, higher IS and lower KL, FD, and FAD indicate better sample quality.}
    \label{fig:evaluation}
\vspace{-4mm}
\end{figure}

The evaluation is performed between a set of generated audio and a set of target audio files. For model evaluation, we follow the metrics used by AudioLDM, including Fréchet Distance~(FD), Inception Score~(IS), Fréchet Audio Distance~(FAD), and Kullback–Leibler~(KL) divergence. All four metrics are calculated based on logits or embedding from audio classifiers. Specifically, IS calculates the entropy of label distribution, where a higher IS indicates a larger variety with vast distinction. KL divergence measures the similarity between generated and target audio by comparing the logits distributions. FAD first computes the multivariate Gaussian of two embedding values collected from a pre-trained VGGish~\cite{vggish}. Then, this score calculates the Fréchet distance between the Gaussian mean and variance. Both KL and FAD indicate better fidelity with lower scores. Besides the three common practices~(IS, FAD and KL) used in previous works~\cite{sampleRNN,audiogen,diffsound}, we also adopt FD, which has a similar idea with FAD but uses PANNs~\cite{panns}, a pre-trained audio pattern recognition model, as the backbone classifier for feature embedding. To compare the effectiveness of these metrics, we perform evaluations between a set of audio files and their corrupted version by the following:

\noindent (1) \textit{Adding noise and random masking.} We add Gaussian noise and mask some information on the mel-spectrogram domain. For Gaussian noise, the mean is the mean value of the mel-spectrogram and the variance equals to $20\%$ of the value range. For masking, we randomly select two places and mask the value of a $10\%$ length of overall mel-spectrogram~(e.g., setting a length of 86 into zero for a $860$ long mel-spectrogram). As Figure~\ref{fig:evaluation} shows, all the metrics can detect this change with a rapid fall or rise.  

\noindent (2) \textit{Adding interference sound.} We randomly select ten irrelevant classes of audio clips and mix them directly with the target sound under the same SNR on mel-spectrogram to verify whether these interfered sounds can be detected. As shown in Figure~\ref{fig:evaluation}, it is obvious that KL and IS do not present significant changes. This might be because adding interfering sounds does not lead to a distinct change in the sound quality. In comparison, FD and FAD can effectively detect changes with an apparent increase in scores.

\noindent (3) \textit{Changing order.} We testify to the sensitivity of these metrics when acoustic events are placed in the wrong order. To simulate this change, the ground-truth data is composed of a group of different sound events and we randomly change their orders. Figure~\ref{fig:evaluation} shows that with the increase of disordered events, only FD presents an increasing trend while other metrics stay stable with little fluctuations. 
Based on the qualitative findings, all four metrics detect the noise efficiently, while only the FD score is capable on classifying irrelevant sounds. 

\begin{table}[htbp]
\caption{The baseline of four datasets on pre-trained AudioLDM}
\centering
\small
\begin{tabular}{cccccc}
\toprule
Dataset                    & Test Condition & FD$\downarrow$ & IS$\uparrow$ & KL$\downarrow$ & FAD$\downarrow$ \\
\midrule
\multirow{2}{*}{ESC50}     & Text      &  $60.63$  &  $5.55$  &  $3.01$  &  $5.95$   \\
                           & Audio     &  $47.46$  &  $6.68$  &  $2.08$  & $4.81$    \\
\midrule
\multirow{2}{*}{US8K}      & Text      &  $31.20$  &  $3.88$  &  $2.20$  &  $10.00$   \\
                           & Audio     &  $32.79$  &  $4.04$   &  $1.44$   &   $13.74$  \\
\midrule
\multirow{2}{*}{AudioCaps} & Text      & $23.63$   &  $6.68$  &  $2.36$  & $4.94$    \\
                           & Audio    &  $21.37$  &  $6.65$  &  $1.78$  &  $2.18$   \\
\midrule
\multirow{2}{*}{AudioSet}  & Text      &  $20.30$  &  $7.56$  &  $2.34$  &   $4.26$  \\
                           & Audio     &  $19.04$  &  $6.72$  &  $1.63$  &  $1.52$  \\
\midrule
\end{tabular}
\label{tab:baseline}
\vspace{-2mm}
\end{table}

\subsection{Benchmark Study}
As shown in Table~\ref{tab:baseline}, we evaluate the performance of pre-trained AudioLDM\footnote{\url{https://github.com/haoheliu/AudioLDM}} as our baselines for text-to-sound generations. Note that we do not perform any fine-tuning on AudioLDM in this section. Although the open-sourced version of AudioLDM is trained with audio embeddings, AudioLDM can perform sampling with either audio or text embedding. Table~\ref{tab:baseline} shows the effect of conditions on different modalities, where AudioLDM conditioned with audio embedding performs better than text embedding in most cases. 
This indicates the distribution of audio and text embedding is not completely aligned, and audio embedding is a more precise conditioning signal for sound generation.  

\begin{table*}[htbp]
\caption{The comparison between different pre-trained strategies. Experiments without pre-training involve building new models from scratch. Audio and text indicate whether the model is taking audio embedding or text embedding as the condition in training. All the results show the best score during training.}
\centering
\small
\begin{tabular}{ccccccccc}
\toprule
 Dataset  & Pre-training & Train Condition & Train Steps~(K) & FD $\downarrow$ & IS $\uparrow$ & KL $\downarrow$ & FAD $\downarrow$\\
\midrule
\multirow{5}{*}{ESC50} 
                       &\XSolidBrush&Audio & $240$      &    $44.75$    &  $7.44$   &  $3.31$   & $4.02$ \\
                       &\XSolidBrush&Text&  $160$     &   $30.74$     &  $10.22$   & $1.84$   & $3.28$ \\
                       & \Checkmark & Audio &   $180$    &   $36.43$     &  $11.15$  &  $2.15$   & $4.41$ \\
                       & \Checkmark & Text &  $80$  & $\mathbf{22.38}$ & $\mathbf{12.98}$ & $\mathbf{1.56}$  & $\mathbf{2.66}$ \\
\midrule
\multirow{5}{*}{US8K} 
                       &\XSolidBrush&Audio& $160$     &   $33.69$     &  $3.73$   &  $2.04$  & $5.75$ \\
                       &\XSolidBrush &Text&    $350$   &    $28.45$    &  $\mathbf{5.00}$   &  $\mathbf{1.87}$  & $\mathbf{4.45}$ \\
                       & \Checkmark & Audio&  $20$     &   $31.21$     &   $3.84$  &  $2.11$  & $7.39$ \\
                       & \Checkmark & Text&   $240$    &    $\mathbf{28.44}$    &  $4.91$  &  $1.88$   & $4.88$ \\
\midrule
\multirow{5}{*}{AudioCaps}
                       &\XSolidBrush &Audio&   $480$   &  $24.04$   &  $7.12$   &   $2.20$    & $2.98$ \\
                       &\XSolidBrush &Text&   $480$     &  $24.84$    &  $6.91$   &  $2.25$     & $2.47$ \\
                       & \Checkmark & Audio&  $80$     &  $\mathbf{23.57}$   &  $7.21$   &   $\mathbf{2.09}$    & $2.98$ \\
                       & \Checkmark & Text&   $240$     &  $25.78$   &  $\mathbf{7.95}$   &   $2.26$    & $\mathbf{1.67}$ \\
\bottomrule
\end{tabular}
\label{tab:pre-train}
\end{table*}

\subsection{Fine-tuning Study}
To study the effectiveness of the pre-trained audio generative model, we fine-tune and evaluate the pre-trained AudioLDM on three smaller datasets, US8K, ESC50 and AudioCaps. Author of~\cite{audioldm} finds audio embedding is better than text embedding in some cases as model condition information. To validate this conclusion in more datasets, we adopt a similar experiment setting and fine-tune AudioLDM with both text embedding and audio embedding as conditioning information. Different from Table~\ref{tab:baseline}, all the experiments in this section only sample with text embedding as input condition since we mainly focus on text-to-sound generation. During the fine-tuning process, we freeze the parameter of CLAP and the VAE encoder, leaving only the latent diffusion model for training. To validate the effect of model pre-training, we also train and evaluate AudioLDM on different datasets from scratch. 

Table~\ref{tab:pre-train} shows the experiment results of this fine-tuning study. We notice that the pre-trained AudioLDM is more advantageous than the model trained from scratch in most cases. With only $32$ samples in each class, the performance of ESC50 can be significantly improved with pre-training. On US8K, the performance of pre-trained AudioLDM is slightly lower, which might attribute to 1) US8K is large enough for model optimization, with around $800$ samples for each class; 2) US8K only contains $10$ sound classes while the pre-trained AudioLDM is capable of generating sound with more diversity, which might degrade model performance on US8K evaluation set. 
Additionally, the pre-trained model can improve generation quality on AudioCaps, particularly on the FAD scores. We also notice that fine-tuning with text embedding on AudioCaps can further achieve a better IS score. We analyze the reason as text embeddings provide weaker conditions, leading to results with less restriction and more diversity. 

AudioLDM is trained in a self-supervised way using audio embedding as conditioning information because training data can be easily scaled up with this scheme. AudioLDM also found that taking audio embedding as the training condition is better than text embedding. However, our experiment shows this is not always the case on different datasets. As shown in Table~\ref{tab:pre-train}, results on small-scale datasets are usually better with text embedding. We hypothesise this is because insufficient audio training data leads to sub-optimal learning of generative models, such as overfitting. This hypothesis is supported by our result, which shows that training models with audio embedding achieve the best performance with fewer training steps, such as $20$k steps in US8K and $80$k steps in AudioCaps.
Conversely, texts or labels provide less detailed and diverse conditions, which can regularize the model to learn data distribution with less chance of overfitting, leading to model convergence with more training steps at the same time.  

\begin{figure}
\centering
\includegraphics[width=1.0\linewidth]{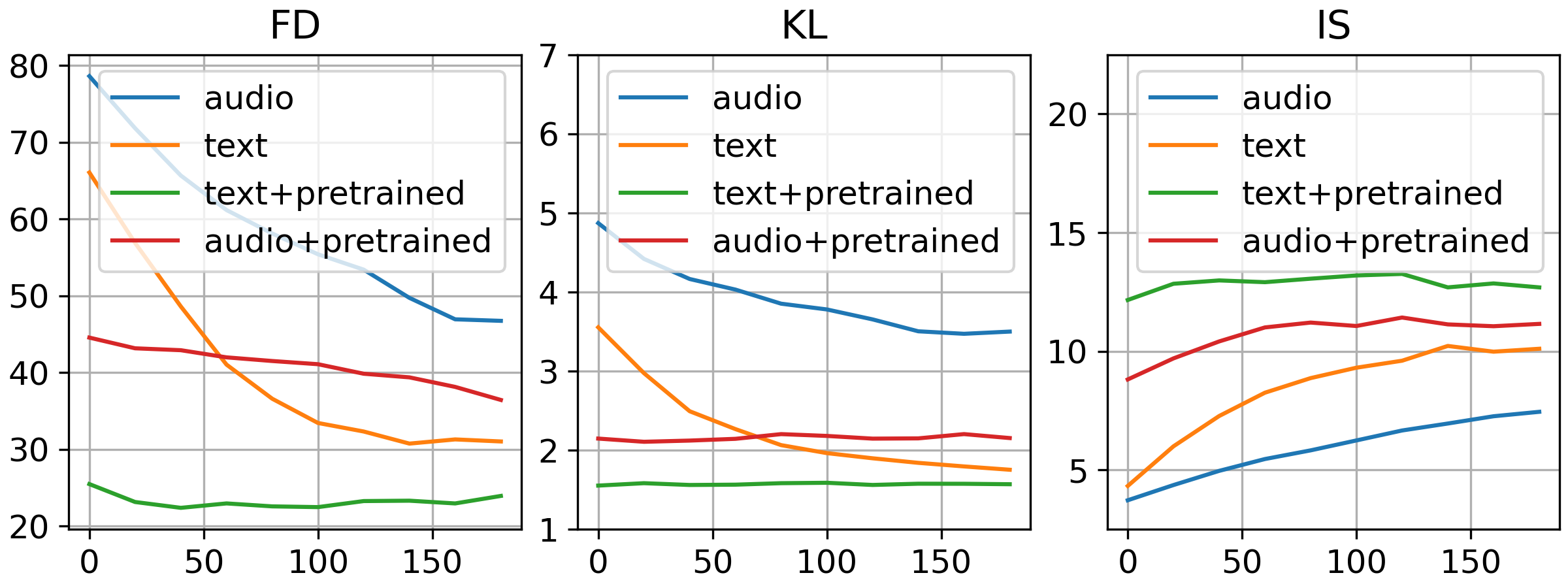}
\caption{The performance of AudioLDM on ESC50 as a function of thousand training steps. Four curves show AudioLDM optimized with 1) audio embeddings; 2) text embeddings; 3) text embedding with pre-trained parameters; and 4) audio embedding with pre-trained parameters }
\label{fig:training_steps}
\vspace{-4mm}
\end{figure}

Figure~\ref{fig:training_steps} illustrates the performance of AudioLDM on different training epochs with and without pertaining and different modalities as condition information. The experiment is performed on the ESC50 dataset. We notice that 1) the pre-trained model can reach coverage quickly with text-embedding, with about \num{20}k training steps; 2) AudioLDM can achieve better performance with text-embedding on ESC50; 3) AudioLDM trained from scratch converge much slower and may not converge well, even with a larger number of steps. 

\section{Conclusion}
This work investigates the effect of pre-trained AudioLDM on sound generation tasks, with results on various settings and datasets. This study shows the pre-trained audio generative model can improve the sample quality and reduce the training time, especially with smaller-scale datasets. This serves as evidence for future studies on audio generation in data-scarcity scenarios. Besides, we conclude that text embedding is preferred as the condition information on small-scale datasets, with the effect of alleviating overfitting during training. Finally, a new benchmark is established for sound generation tasks with four commonly used datasets. These baseline results can be used as a reference for future studies of sound generation.

\section{Acknowledgment}
This research was partly supported by a research scholarship from the China Scholarship Council~(CSC) No.$202208060240$, the British Broadcasting Corporation Research and Development~(BBC R\&D), Engineering and Physical Sciences Research Council~(EPSRC) Grant EP/T019751/1 ``AI for Sound'', and a PhD scholarship from the Centre for Vision, Speech and Signal Processing~(CVSSP), University of Surrey. For the purpose of open access, the authors have applied a Creative Commons Attribution~(CC BY) license to any Author Accepted Manuscript version arising.


\end{document}